\documentclass[letter]{aa}

\usepackage{times}
\usepackage{footmisc}
\usepackage{graphicx,multicol}
\usepackage{caption}
\usepackage{subcaption}
\usepackage{natbib}

\usepackage{amsmath,amsfonts,bm}
\usepackage{amsthm}
\bibpunct{(}{)}{;}{a}{}{,} 
\usepackage{txfonts}
\usepackage{relsize}
\usepackage{soul}
\usepackage[draft=False, colorlinks=true, citecolor=MidnightBlue, urlcolor=MidnightBlue, linkcolor=MidnightBlue, linktoc=all]{hyperref}
\usepackage{cleveref}
\usepackage[section]{placeins}
\usepackage[switch, mathlines]{lineno}

\newcommand\facc{$f_\mathrm{acc}~$}
\newcommand\gfacc{$\mathrm{^{global}}f_\mathrm{acc}(t)~$}

\begin{document}

\title{Turn up the light: Radiative efficiency of protostars at birth}

\author{Asmita~Bhandare\inst{\ref{lmu},\ref{lyon}}
\fnmsep \thanks{\href{asmita.bhandare@lmu.de}{asmita.bhandare@lmu.de}}\textsuperscript{$\ddagger$}\orcid{0000-0002-1197-3946}
\and Adnan Ali Ahmad \inst{\ref{lyon}}
\fnmsep \thanks{\href{adnan-ali.ahmad@cnrs.fr}{adnan-ali.ahmad@cnrs.fr}}\thanks{Both authors have contributed equally to this work.}\orcid{0009-0004-2719-1107}
\and Beno\^{i}t~Commer\c{c}on\inst{\ref{lyon}}
\orcid{0000-0003-2407-1025}
}

\institute{
Universit\"ats-Sternwarte, Fakult\"at f\"ur Physik, Ludwig-Maximilians-Universit\"at M\"unchen, Scheinerstr.~1, 81679 M\"unchen, Germany
\hspace{-1em} 
\label{lmu}
\and
ENS de Lyon, CRAL UMR5574, Universite Claude Bernard Lyon 1, CNRS, Lyon, 69007, France
\label{lyon}
}

\date{Submitted: June 30, 2025; Accepted: September 8, 2025}

\abstract{ 
Early stages of stellar birth comprise of a two-step process involving the formation of two hydrostatic cores. The second step of gravitational collapse sets the radiative efficiency and accretion rate of the young protostar. These two parameters, of prime importance for protostellar evolution, dictate the luminosities and thus play a key role in deciphering the current discrepancy between observational surveys and theoretical models. In this letter, we provide quantitative estimates on the evolution of the radiative efficiency and accretion rate obtained from self-consistent, high-resolution, radiative hydrodynamic simulations performed using the codes \texttt{PLUTO} and \texttt{RAMSES}. The main highlight of our result is that the radiative efficiency reaches unity, that is, supercriticality, relatively quickly after protostellar birth. Supercriticality at the accretion shock is a necessary condition for cold accretion. Our results thus support a rapid transition to the cold accretion scenario, which is one of the assumptions used in Pre-Main Sequence (PMS) models working towards solutions to explain observational data. We briefly discuss the implications of the time evolution of the radiative efficiency factor in the context of the luminosity problem, the Protostellar Luminosity Function (PLF), PMS evolution, accurate sink properties, and the stellar Initial Mass Function (IMF). }


\authorrunning{Bhandare, Ahmad \& Commer\c{c}on}

\maketitle

\graphicspath{{./Plots/}}

\section{Introduction}
\label{sec:intro}

Understanding the birth of stars through the gravitational collapse to the onset of hydrogen fusion is a challenging task: modelling the birth of a protostar requires one to describe a highly non-linear and complex process that spans orders of magnitude in both density and spatial extent. Furthermore, modelling protostellar evolution following its formation relies on Pre-Main sequence (PMS) theory. A crucial hypothesis in this theory is that of cold accretion \citep{palla_1993, Hartmann_1997, hosokawa_2009, Baraffe2017}. In PMS models this is expressed by the factor of ($1 - \alpha$) where $\alpha$ represents the fraction of accretion energy absorbed at the protostellar surface and cold accretion implies $\alpha=0$. A number of processes could explain a drop in $\alpha$, which may be mechanical in nature, such as with outflows, jets, and boundary disc accretion, or radiation. In this letter, we draw a parallel between said $\alpha$ parameter and the radiative efficiency of the protostellar accretion shock $f_{\mathrm{acc}}$, which represents the fraction of the incident accretion energy radiated away. $f_{\mathrm{acc}}=1$ is a necessary condition for cold accretion.

Recent numerical studies on protostellar collapse including radiation hydrodynamics have offered insights on the radiative properties of the accretion shocks on the first and second hydrostatic cores. Doing-so requires a self-consistent modelling of the two shock fronts in which one describes protostellar formation following the second gravitational collapse. This was pioneered by \cite{Larson1969}, who discovered the two-step evolutionary sequence that results in the birth of a protostar deeply embedded in its surrounding envelope (see Fig.~\ref{fig:collapse}). At first, the collapse of the gravitationally unstable dense molecular cloud core proceeds isothermally, as the gravitational binding energy is efficiently radiated away by dust grains as infrared thermal emission. The resulting short cooling timescale enables the gas to maintain thermal equilibrium with the surrounding interstellar radiation field, typically $\sim$10-30~K. In the central region, gas density and hence optical depth increases due to gravitational compression. This causes the heating rate to exceed the cooling rate. Thus, the collapse enters an almost adiabatic phase where the gas further heats up. This builds up thermal pressure support leading to a hydrostatic equilibrium, forming the first core. The first core mass increases as it continues accreting material, maintaining a quasi-hydrostatic equilibrium while slowly contracting adiabatically, until temperatures exceed $\approx$2000 K, by which point molecular hydrogen (H$_{2}$) dissociation begins. This endothermic process extracts heat from the gas to dissociate H$_{2}$, which was otherwise being injected into the thermal energy budget of the first core. As a result, the polytropic index of the gas, previously at 7/5, drops to $\approx$1.1, below the critical value of 4/3 needed for stability against gravitational collapse. This causes a violent second collapse within the first core, giving birth to the second core (i.e., the protostar) once all H$_2$ molecules are dissociated.

Since these calculations are complex and expensive to run, only a handful studies have investigated the radiative behaviour of the protostellar accretion shock (see the review by \citealt{Teyssier2019}). It is however crucial to note that multi-dimensional calculations resolving the protostar are often stopped soon after protostellar birth \citep[e.g.,][]{Tomida2013, Vaytet2018} owing to stringent time-stepping constraints, with only a few studies following its evolution for a few tens of years afterward \citep[e.g.,][]{Bate2010, Schonke2011, machida_2019, wurster_2022}.

Radiative shocks with a radiative efficiency \mbox{\facc = 1} are commonly referred to as \textit{supercritical}, whereas with \mbox{\facc < 1} are referred to as \textit{subcritical} \citep[see][]{zeldovich1967}. The first core's accretion shock has been shown to be supercritical \citep{Commercon2011, Vaytet2013, Bate2014, Vaytet2018, Bhandare2018}; in contrast, the protostar's shock front at birth seems to be strongly subcritical \citep{Vaytet2013, Bate2014, Vaytet2018, Bhandare2020, Ahmad2023}. The swelling of the protostar to very large radii immediately after its birth, followed by a contraction phase, was already reported in older uni-dimensional spherically symmetrical calculations in the literature, having studied the collapse across longer timescales \citep[e.g.,][]{Larson1969, narita_1970, winkler_1980, Stahler1980a, Stahler1980b, masunaga_2000, Vaytet2017}. This seems to imply that the protostellar accretion shock is initially subcritical and later exhibits a transition to supercriticality, as the addition of significant amounts of high entropy material to the protostellar interior causes it to swell. However the precise value of $f_{\mathrm{acc}}$, that is, the fraction of the accreted gravitational energy radiated away at the shock, and its evolution with time remained unexplored. 

This radiative efficiency and the early accretion history of protostars has been used in solutions to explain the observed low and widely spread protostellar luminosities \citep[see reviews by][]{dunham_2014, fischer_2023}. It is also used as a free parameter to compute the radiative feedback of sink particles and from sink cells, a replacement for protostars in simulations \citep[e.g.,][]{Offner2009, Commercon2022}, with further implications on constraining the mass distribution of stars in the stellar Initial Mass Function \citep[IMF;][]{Hennebelle2020, Hennebelle2022}. 

In this letter, we provide a quantitative estimate of the radiative efficiency of the protostellar accretion shock using numerical experiments and how it varies over time after protostellar birth. To do so, we present measurements of $f_{\mathrm{acc}}$\footnote{Note that this is not the \textit{effective} $f_{\mathrm{acc}}$, which also includes factors due to outbursts and non-radiative losses (via outflow/jet/wind), for example, considered in \citet{Dunham2010, Offner2011}.} and accretion rates from second collapse calculations performed independently using two radiation hydrodynamic codes \texttt{PLUTO} and \texttt{RAMSES}. Three-dimensional (3D) simulations by \citet{Vaytet2018} and \citet{Ahmad2024} show that the second core radiates preferentially along the poles, in the direction of least optical depth. This indicates that multi-dimension is key for a complete picture. Hence, the results are validated by comparing one- and two-dimensional simulation outputs from \citet{Bhandare2018, Bhandare2020, Bhandare2024} with 3D calculations from \citet{Ahmad2024}. 

The letter is organised as follows. In Sect.~\ref{sec:results}, we focus on the evolution of the radiative efficiency and accretion rate over the initial period of the protostellar lifetime and the implications of these results are discussed in Sect.~\ref{sec:discussion}. We summarise the main takeaways in Sect.~\ref{sec:summary}.

\section{Results}
\label{sec:results}

In order to provide a quantitative measurement of the radiative efficiency of the protostellar shock front, we carry out fully self-consistent\footnote{No sink particle or cell is used in these simulations, and the protostar's hydrostatic equilibrium is resolved.} radiation hydrodynamic calculations, describing the birth of the protostar from the gravitational collapse of an unstable molecular cloud core. We present the results of spherically symmetrical simulations from \citet{Bhandare2018, Bhandare2020} performed using the \texttt{PLUTO} code. Additionally, we compare these with results from calculations that include an initial angular momentum in the molecular cloud core with the \texttt{PLUTO} and \texttt{RAMSES} codes discussed in \citet{Bhandare2024} and \citet{Ahmad2024}, respectively. The technical details and simulation set-ups used are briefly presented in Appendix~\ref{appendix:methoddetails}. 

Quantitative descriptions of the properties of the first and second hydrostatic cores can be found in \citet{Bhandare2018, Bhandare2020} and \citet{Ahmad2023, Ahmad2024}. Here we focus on the fraction of accretion energy radiated away, that is, the radiative efficiency of the protostellar shock front \facc defined as
\begin{align}
\mbox{${f}_\mathrm{acc} \equiv \dfrac{F_\mathrm{rad}}{F_\mathrm{acc}}$},
\end{align} 
with the radiative flux in the FLD approximation computed as
\begin{align}
\mbox{$ F_\mathrm{rad} = - ~\dfrac{\lambda c}{\kappa_\mathrm{R} \rho}  ~\nabla E_\mathrm{rad}$},
\label{eq:flux}
\end{align}
where $c$ is the speed of light, $\kappa_\mathrm{R}$ is the Rosseland mean opacity, $\rho$ is the gas density, and $E_\mathrm{rad}$ is the radiative energy. The flux limiter $\lambda$ is chosen following \citet{Levermore1981} for the \texttt{PLUTO} runs and \citet{Minerbo1978} for the \texttt{RAMSES} runs. The limiting cases of diffusion and free streaming are recovered by the flux limiter. Following \citet{Ahmad2024}, the accretion energy flux $F_\mathrm{acc}$ is computed as
\begin{align}
\mbox{$F_\mathrm{acc} \simeq -\rho v_r \dfrac{G M_\mathrm{sc}}{R_\mathrm{sc}} -E_\mathrm{int}v_r -Pv_r -\dfrac{\rho v_r^3}{2} $},  
\label{eq:accflux}
\end{align}
where $v_r$ is the radial velocity, $G$ is the gravitational constant, $M_\mathrm{sc}$ is the mass enclosed within the second core radius $R_\mathrm{sc}$, $E_\mathrm{int}$ is the internal energy, and $P$ is the gas pressure. The right-hand side of Eq.~\ref{eq:accflux} consists of the gravitational potential energy flux, the internal energy flux, pressure contribution, and the kinetic energy flux. Among these, the first and last terms are dominant.

Both $F_\mathrm{rad}$ and $F_\mathrm{acc}$ are estimated just outside the second core radius, meaning that we perform a local measurement of $f_{\mathrm{acc}}$. This approach is consistent with PMS models that study the protostar's thermal adjustment according to radiative conditions at the shock front. $F_\mathrm{rad}$ includes both the internal cooling flux\footnote{Blackbody radiation emitted by the gas downstream of the accretion shock (i.e., from within the protostellar interior).} from the second core and the accretion luminosity released at the accretion shock. However, the interior flux is negligible when compared to that produced at the shock front as we are modelling the earliest phases of star formation. In the 3D \texttt{RAMSES} run, the measurement is performed along the polar regions, where the protostar shines most brightly. Two rays are launched along the north and south poles (defined by the angular momentum axis of the protostar), and an average of $f_\mathrm{acc}$ is performed\footnote{See also Appendix~\ref{appendix:globalfacc} for a global measurement of $f_\mathrm{acc}$ in this run.}.

We compare \facc against the accretion rate onto the second core
\mbox{$ \dot{M}_\mathrm{sc} = 4 \pi ~{R_\mathrm{sc}^2} ~\rho_\mathrm{sc} ~|v_\mathrm{sc}| $},
where $\rho_\mathrm{sc}$ and |$v_\mathrm{sc}$| are the density and absolute value of radial velocity at the second core radius $R_\mathrm{sc}$, defined to be the location of the second accretion shock. 

\begin{figure}[!tp]
\centering
\begin{subfigure}{0.45\textwidth}
\includegraphics[width=\textwidth]{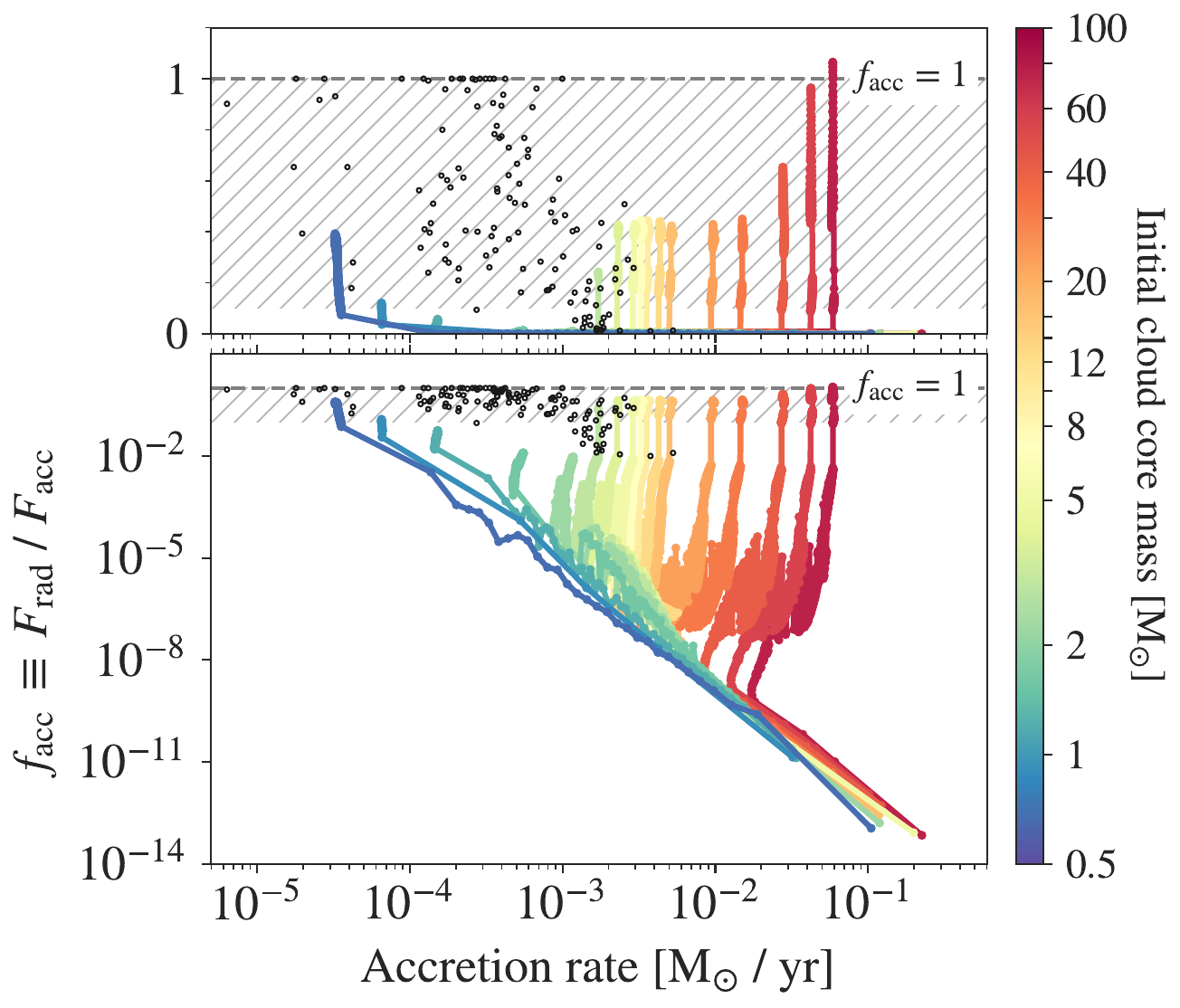}
\end{subfigure}
\begin{subfigure}{0.45\textwidth} \includegraphics[width=\linewidth]{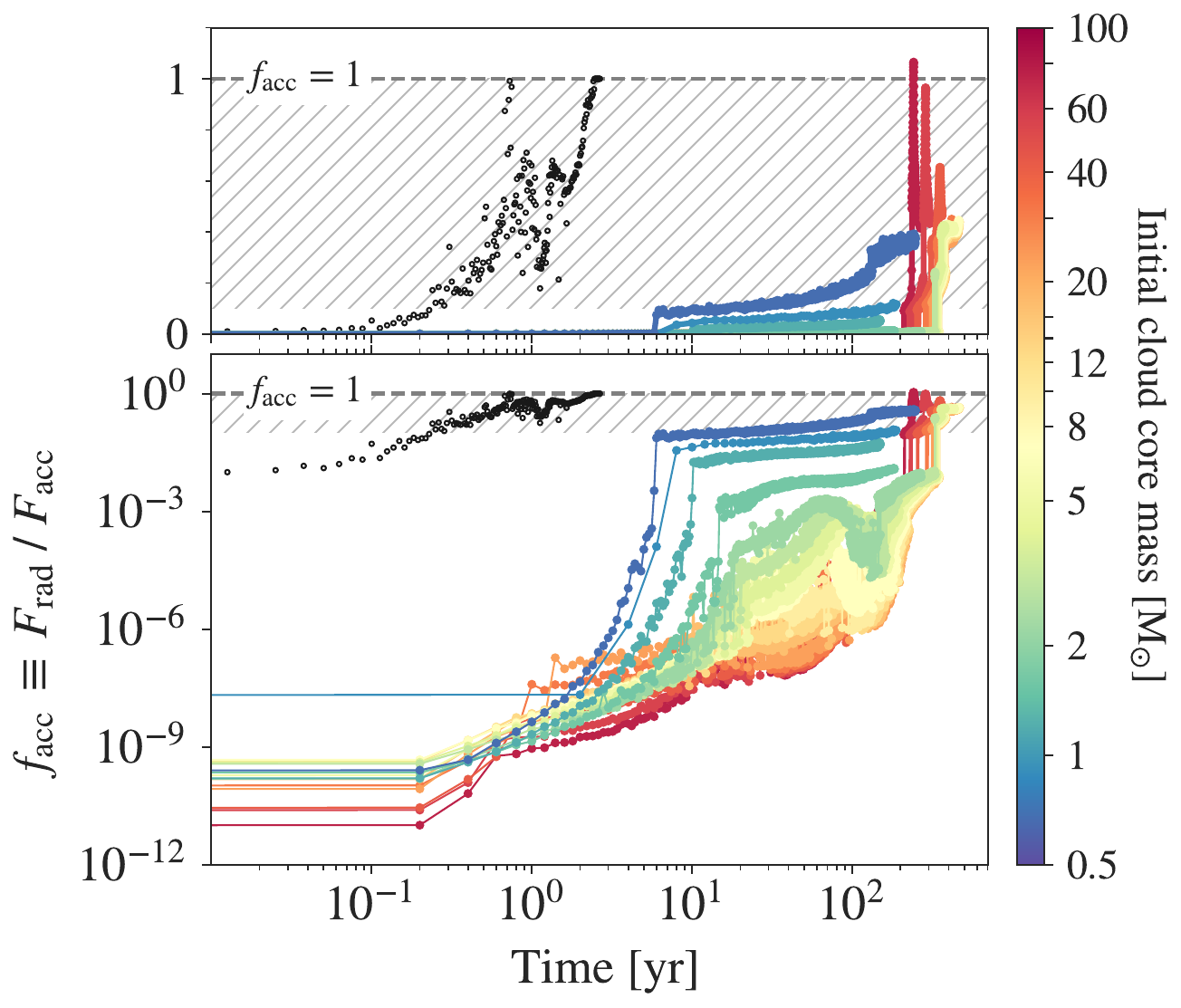}
\end{subfigure}
\caption{Radiative efficiency \facc versus mass accretion rate $\dot{M}_\mathrm{sc}$ at the protostellar surface (\textit{top}) and time since protostellar birth (\textit{bottom}). Both panels show outcomes from a 1D spherical set-up with different initial molecular cloud core masses using the \texttt{PLUTO} code. Also shown as black circles in both panels is data from a 1~$M_{\odot}$ 3D \texttt{RAMSES} simulation with an initial turbulent Mach number of 0.2. Each plot includes a top subplot with a linear \facc axis ranging from 0 to 1. Typical constant values of \facc used in the literature lie within the grey hatched region.}
\label{fig:Avg-outsideshock}
\end{figure}

In this main text, we focus on 1D spherically symmetric \texttt{PLUTO} runs and a 3D \texttt{RAMSES} run including an initial angular momentum. The 1D \texttt{PLUTO} and 3D \texttt{RAMSES} data for the time evolution of the radiative efficiency and accretion rate are shown in Fig.~\ref{fig:Avg-outsideshock}. The time evolution of $F_\mathrm{rad}$ and $F_\mathrm{acc}$ are independently shown in Appendix~\ref{appendix:localfacc}. For completeness and comparison, we refer the reader to Fig.~\ref{fig:2DAvg-outsideshock} in Appendix~\ref{appendix:2DPluto} for the 2D \texttt{PLUTO} results with and without an initial angular momentum. We follow the protostellar evolution for $>$100~years after its formation until the central density reaches $\approx$~0.5--0.8 g~cm$^{-3}$ in the 1D spherically symmetric \texttt{PLUTO} runs. The computational constraints of a full 3D run are such that long time integrations are prohibitively expensive, especially when resolving the second core. Hence, the 3D \texttt{RAMSES} simulation data is limited to $\sim$2.5 years of protostellar evolution.

The initially high accretion rate gradually decreases as the first core is accreted onto the protostar, followed by an almost steady accretion of the surrounding envelope. The accretion shock onto the new-born protostar is strongly \textit{subcritical} (\mbox{\facc < 1}), meaning that most of the incident kinetic energy is transferred to the protostar. This huge energy and mass input results in a ``bloating'' phase\footnote{The Kelvin–Helmholtz timescale is shown to largely exceed the accretion timescale during this bloating phase in \citet{Bhandare2020}.}, that is, a fast expansion of the protostellar shock front, used to define the protostellar radius. As seen in our simulations, protostars undergo an initial expansion phase right after birth where \facc is much lower. While the radiative efficiency of the protostar stays well below unity during very early evolution, it rapidly ($<$100~years) increases. In the high-mass end of the 1D simulations, as \facc reaches unity comparatively faster than for the lower-mass cases with a similar central density, the protostar contracts once it can start radiating away the energy accumulated over a few tens of years after its birth. In the 1~$M_{\odot}$ 3D run, \facc reaches unity much earlier during protostellar evolution since the second core with its surrounding disc radiates preferentially in the direction of least optical depth, that is, along the pole. The 1D simulation results thus represent a limiting case where the system struggles to efficiently radiate away accretion energy. The rise in \facc occurs differently in 1D and 3D models. In 1D, the shock front expands outward, with \facc showing a rapid increase (set by the first core's free-fall time) followed by a plateau as the cloud core remnants are slowly accreted (Fig.~\ref{fig:Avg-outsideshock}). In 3D, $f_{\mathrm{acc}}=1$ results much earlier from geometry: the polar cavities allow efficient radiation of the accretion energy. The implications of this transition from sub- to supercriticality very early on in the protostellar lifetime are discussed in the next section.

\section{Discussion}
\label{sec:discussion}

In this section we discuss the significance of \facc from our collapse calculations in the wider context of the field; namely, how the result fits into the luminosity problem and PMS evolution, as well as how it aids in better constraining sink properties and the stellar IMF.

Observations of protostellar luminosities and accretion rates have consistently fallen below theoretical predictions \citep{Kenyon1990, Kenyon1995}, creating the protostellar luminosity problem. Improved measurements (better sampling, revised star formation timescales, and extinction corrections) have reduced this discrepancy \citep{Evans_2009, Enoch2009, Dunham2010, Dunham2013, Fischer_2017}. However, newer surveys reveal a spread spanning orders of magnitude in the Protostellar Luminosity Function (PLF; the modern luminosity problem\footnote{Not to be confused with the `classical' luminosity problem.}, now under active study, \citealp[see][]{fischer_2023}). To explain the large scatter in the PLF, studies have explored prolonged star formation (extended duration of reduced accretion) and episodic accretion (bursts from disc instabilities, fragmentation, outflows, or stellar encounters), as the accretion luminosity depends on the accretion rate. For reviews, see \citet{dunham_2014} and \citet{fischer_2023}. Protostellar accretion can also alter the thermal structure, shortening the lithium depletion timescale, explaining the observed spread in surface lithium abundance in low-mass PMS stars \citep[see][]{Tognelli2021}. 

The accretion luminosity also depends on the \facc factor, which represents the amount of incident kinetic energy that is radiated away at the protostellar accretion shock. Models providing a solution to the classical luminosity problem and the spread in the PLF \citep[e.g.,][]{Evans_2009, Dunham2010, Padoan2014, Kospal2017, Fischer_2017} rely on the ad hoc assumption that the protostellar shock front is supercritical (\facc$\approx$~1); however, this is not based on a rigorous analysis of the radiative behaviour of the accretion shock. 

The $\alpha$ parameter significantly impacts PMS evolutionary tracks \citep[e.g.,][]{prialnik_1985, palla_1993, Hartmann_1997, hosokawa_2009, baraffe_2009, Baraffe2012, Baraffe2017, Tognelli2020, amard_2023}. While purely cold accretion is often assumed in PMS models to explain observational data, this may not hold during and shortly after ($<$100~years) protostellar birth. A hybrid scenario, switching from hot ($\alpha \geq$~0.1-0.2) to cold ($\alpha \approx 0$) accretion at a fixed rate threshold, fails to explain the luminosity spread \citep{Hosokawa2011, Baraffe2017}. Instead, the switch based on a non-constant $\alpha$ varying with accretion rate better matches observed PMS tracks and lithium abundance \citep{Hosokawa2011, Baraffe2012, Baraffe2017}. Our results, showing changes in the radiative efficiency \facc and accretion rate, support this idea. Regardless of the mechanisms responsible for lowering $\alpha$ (e.g., outflows, disc accretion), once the accretion shock becomes radiatively supercritical, it rapidly ($<$100~years) radiates residual accretion energy, ensuring a transition to cold accretion. 

The protostar is also often replaced with a sink particle or cell, modelled using sub-grid prescriptions \citep[e.g.,][]{Li2018, Grudic2022}, enabling exploration of larger spatial scales over longer timescales. In such simulations, $f_{\mathrm{acc}}$, treated as a free parameter, contributes to the radiative feedback emitted by the sink, which in turn affects the IMF by shifting its peak and increasing the number of massive stars through reduced fragmentation \citep[e.g.,][]{Hennebelle2020, Hennebelle2022}.

Furthermore, in observational data analyses, \facc is also used to estimate the mass accretion rates onto Class 0-I-II protostars, where it is often assumed to be 100\% \citep[e.g.,][]{Kenyon1990, Enoch2009, Fiorellino2023}. 

Accurate estimates of \facc and the accretion rate over time from self-consistent simulations such as those presented here thus serve as valuable inputs, providing justification for the use of these otherwise free parameters.

\section{Summary}
\label{sec:summary}

In this letter, we provide quantitative estimates for the time-varying radiative efficiency parameter \facc and non-steady accretion rate with the self-consistent modelling of the accretion shock across a wide range of initial cloud core masses. 

Our 1D, 2D, and 3D radiation hydrodynamic simulations show that at birth a significant fraction of accretion energy is advected into the young protostar. The radiative efficiency at the protostellar accretion shock is extremely low immediately following its birth. However, \facc increases significantly in the first few tens of years of evolution, already reaching the current fixed values used in the literature of \facc $\approx$ 0.1 -- 1, when considering mechanical effects \citep{Offner2009, Offner2011, McKee2011, Vorobyov2015, Hennebelle2020}. Thus, very early on there is transition from sub- to supercriticality, which occurs fastest for the higher-mass regime. These results are consistent with a rapid switch after protostellar birth from hot to cold accretion, supporting a hybrid scenario commonly used in PMS evolution models.

\begin{acknowledgements} 
We thank the referee for the detailed feedback, which helped improve the clarity of this letter. We also thank Isabelle Baraffe and Gilles Chabrier for their valuable input on the draft. AB acknowledges funding by the Deutsche Forschungsgemeinschaft (DFG, German Research Foundation) under Germany's Excellence Strategy - EXC-2094 - 390783311. BC and AAA acknowledges funding from the French Agency Nationale de la Recherche (ANR) through the projects DISKBUILD (ANR-20-CE49-0006) and PROMETHEE (ANR-22-CE31-0020). We gratefully acknowledge support from the CBPsmn (PSMN, P\^{o}le Scientifique de Mod\'{e}lisation Num\'{e}rique) of the ENS de Lyon for the computing resources. The platform operates the SIDUS solution developed by Emmanuel Quemener \citep{sidus}.

\end{acknowledgements}

\bibliographystyle{yahapj}

\bibliography{Bibliography}

\begin{appendix}

\section{Protostellar collapse sequence}

\begin{figure}[hb!]
\centering
\begin{subfigure}{0.44\textwidth}
\includegraphics[width=\linewidth]{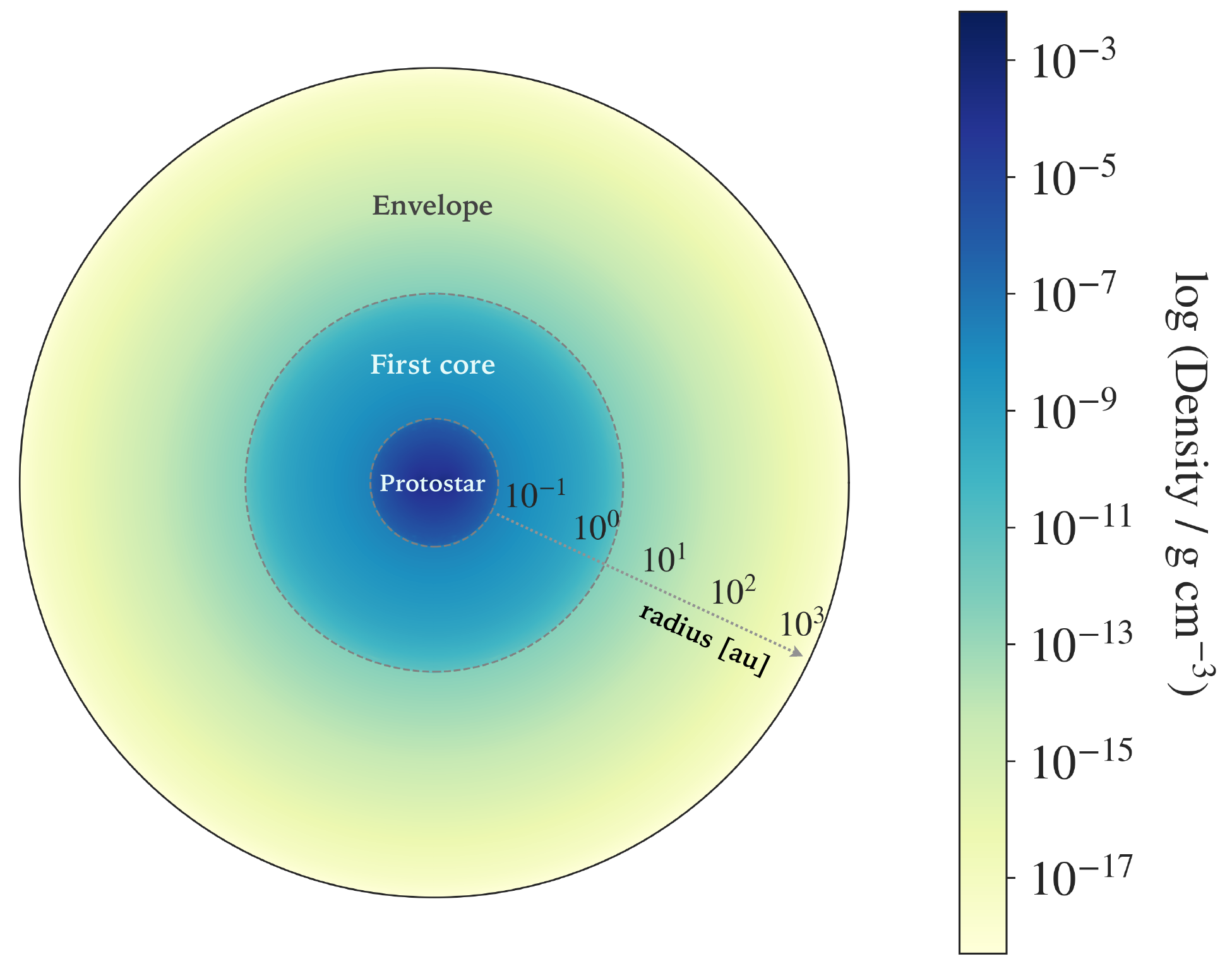}
\vspace{0.2cm}
\end{subfigure}
\begin{subfigure}{0.45\textwidth} 
\includegraphics[width=\linewidth]{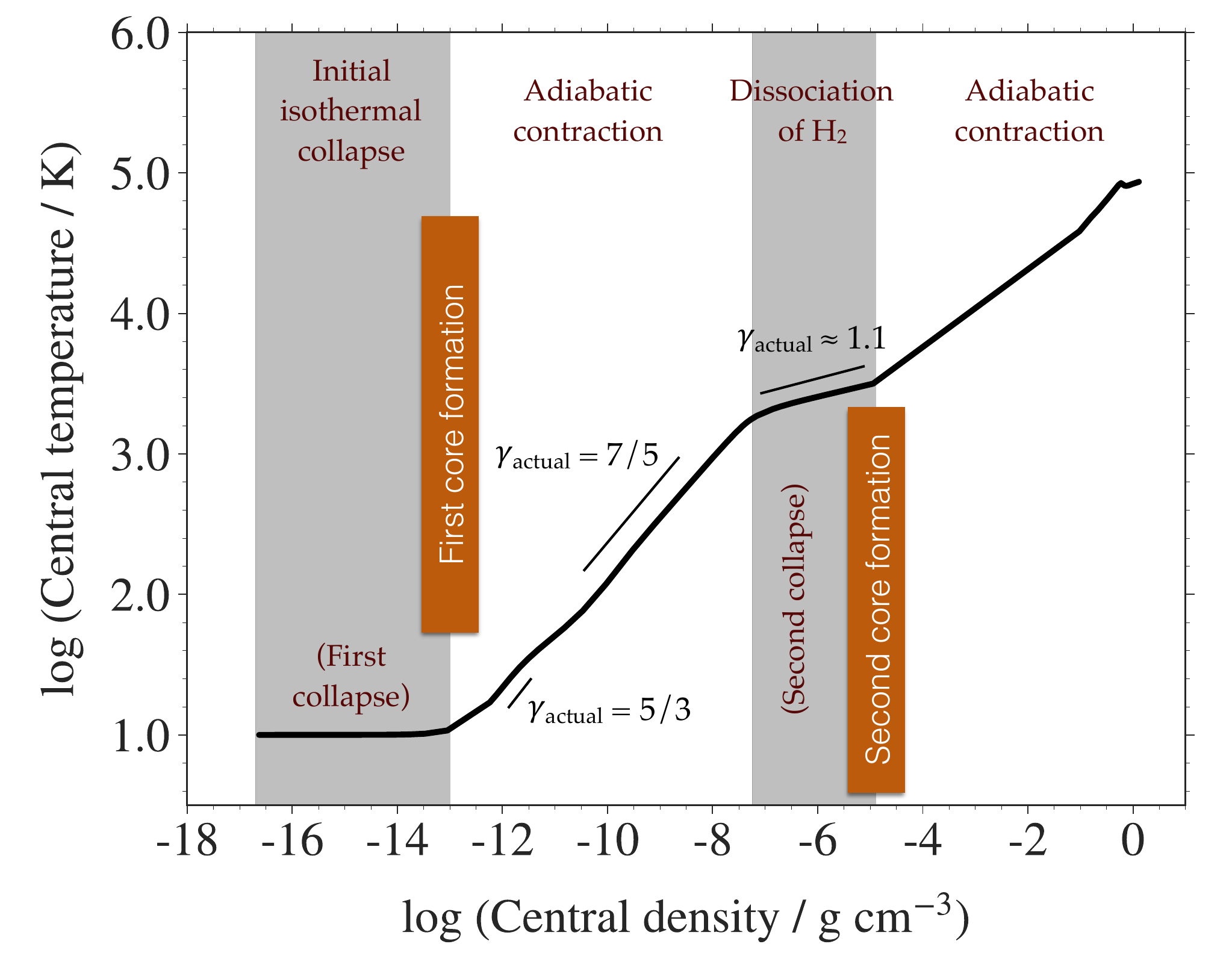}
\end{subfigure}
\caption{\textit{Top:} Logarithmically scaled (size in au) time snapshot showing the gas density in colour within an infalling envelope, the first core at its final stage, and the onset of the protostar (second core) formation. \textit{Bottom:} Thermal evolution of a 1~$M_{\odot}$ molecular cloud core highlighting the two-step collapse process. Bottom panel is reproduced from \citet{Bhandare2020}.}
\label{fig:collapse}
\end{figure}

Figure~\ref{fig:collapse} summarises the two-step gravitational collapse process that leads to protostellar birth. The top panel shows a time snapshot displaying the presence of the outer first hydrostatic core and the inner second core (i.e., protostar), both embedded in the surrounding envelope (i.e., remnant of the molecular cloud core). Further details of this collapse process and properties of the two hydrostatic cores can be found in our previous works \citep[e.g.][]{Bhandare2018, Bhandare2020, Ahmad2023, Ahmad2024}.

\section{Numerics and initial set-up}
\label{appendix:methoddetails}

In this section, we briefly describe the technical details of the radiation hydrodynamic simulations performed with the codes \texttt{PLUTO} and \texttt{RAMSES}. The effects of magnetic fields on protostellar properties will be investigated in future work\footnote{Magnetised 3D runs using a similar set-up are presented in \cite{ahmad_2025}. These have yielded very similar results to the hydrodynamical runs presented in this letter.}.

\subsection{\texttt{PLUTO} simulations}
The simulation data\footnote{\texttt{PLUTO} data is available via \href{https://asmitabhandare.github.io/\#tara}{asmitabhandare.github.io/\#tara}.} used herein is an outcome of one- and two-dimensional (1D, 2D) molecular cloud core collapse calculations performed with a modified version of the radiation hydrodynamic code \texttt{PLUTO} \citep[version 4.1, 4.4;][]{Mignone2007, Mignone2012}. Details of the 1D and 2D radiation hydrodynamic simulation set-up for a spherical collapse can be found in \citet{Bhandare2018, Bhandare2020} and the 2D set-up including an initial angular momentum prescribed as solid-body rotation to enable disc formation is detailed in \citet{Bhandare2024}.

We treat the self-gravity using the \texttt{HAUMEA} \citep{Kuiper2010b, Kuiper2011} module and use the \texttt{MAKEMAKE} \citep{Kuiper2010a} module to account for the radiation transport via a grey (frequency-averaged) flux-limited diffusion (FLD) approximation. A variable gas equation of state from \citet{Dangelo2013} and \citet{Vaidya2015} is used to study the effects of $\mathrm{H_2}$ dissociation and molecular vibrations and rotations. We make use of tabulated dust opacities from \citet{Ossenkopf1994} and gas opacities from \citet{Malygin2014}. 

1D simulations span initial cloud core mass ranging from 0.5 -- 100~$M_{\odot}$. 2D spherical collapse simulations sample a few cases of 1, 5, 10, and 20~$M_{\odot}$, while we focus only on the fiducial 1~$M_{\odot}$ collapse for the run with rotation. All simulation runs start with a Bonnor--Ebert sphere like density profile \citep{Ebert1955, Bonnor1956} and a 10~K temperature. The initial ratio of thermal to gravitational energy varies depending on the initial cloud core mass. As an example, for the 1~$M_{\odot}$ case it is fixed to 0.29. We set a solid-body rotation rate of \mbox{$\Omega_\mathrm{0} = 1.77 \times 10^{-13}$ rad~$\mathrm{s}^{-1}$} to initiate the run with angular momentum. 

The computational grid for the 1D simulations consists of 320 uniformly spaced cells spread between $10^{-4}-10^{-2}$~au and 4096 logarithmically spaced cells across $10^{-2}-3000$~au yielding a smallest cell size of $\Delta x_\mathrm{min} = \Delta r = 3.09\times10^{-5} $~au. For the 2D spherical collapse simulations, 1445 logarithmically spaced cells are used in the radial direction from $10^{-2}-3000$~au and 180 uniformly spaced cells are assigned stretching from the pole ($\theta = 0^\circ$) to the midplane ($\theta = 90^\circ$) with the smallest cell size of $\Delta x_\mathrm{min} = \Delta r = r \Delta\theta = 8.77\times10^{-5}$~au. The 2D simulations including solid-body rotation use the same radial extent with 241 logarithmically spaced cells and the polar extent with 30 uniformly spaced cells resulting in the smallest cell size of $\Delta x_\mathrm{min} = \Delta r = r \Delta\theta = 5.37\times10^{-4}$~au.

\subsection{\texttt{RAMSES} simulations}
We compare the 1D and 2D \texttt{PLUTO} results with a similar set-up in 3D using the \texttt{RAMSES} code \citep{teyssier_2002} described in \citet{Ahmad2023, Ahmad2024}. Radiative transfer is treated using the grey FLD approximation \citep{Commercon2011a, commercon_2014, gonzalez_2015}. The simulations use the \cite{saumon_1995} gas equation of state and the opacity table pieced together by \cite{Vaytet2013}, which describes both dust as well as molecular and atomic gas opacities. The initial conditions consist of a uniform density sphere of 1~$M_{\odot}$, with a temperature of $10$ K, and a thermal to gravitational energy ratio of $0.25$. Here, we present a single \texttt{RAMSES} run containing net angular momentum through the inclusion of a turbulent initial velocity vector field, parametrised by a turbulent Mach number, which has been set to 0.2. The computational grid is refined according to the local Jeans length (computed at a constant temperature of 100 K), and we allow 19 levels of adaptive mesh refinement, yielding a spatial resolution of $\Delta x_{\mathrm{min}} = 2.9\times10^{-4}$ AU at the finest level. Further details of this simulation are mentioned in \citet{Ahmad2024}.

\section{Two-dimensional \texttt{PLUTO} runs}
\label{appendix:2DPluto}

\begin{figure}[!tp]
\centering
\begin{subfigure}{0.4\textwidth} 
\includegraphics[width=\linewidth]{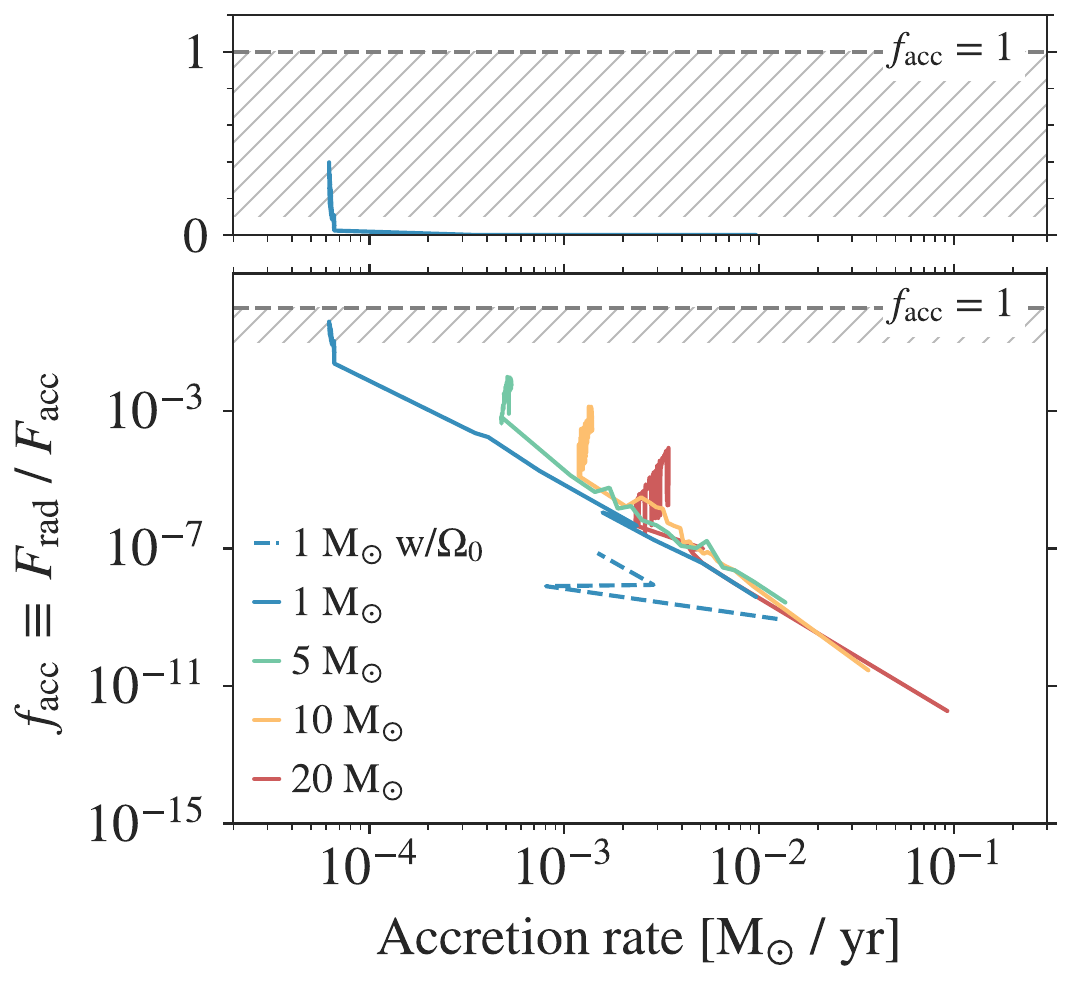}
\vspace{0.1cm}
\end{subfigure}
\begin{subfigure}{0.4\textwidth} 
\includegraphics[width=\linewidth]{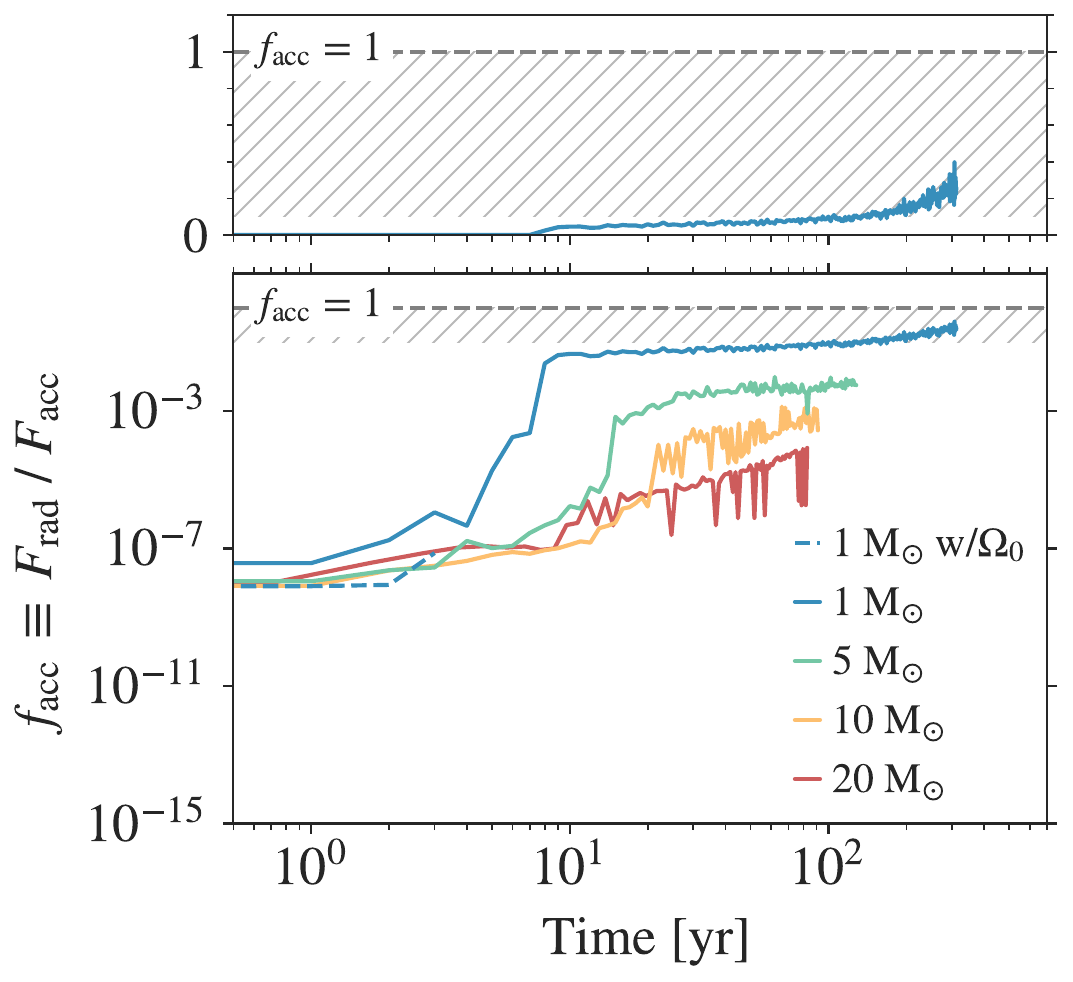}
\end{subfigure}
\caption{Radiative efficiency \facc versus mass accretion rate $\dot{M}_\mathrm{sc}$ at the protostellar surface (\textit{top}) and time since protostellar birth (\textit{bottom}). Both panels show \texttt{PLUTO} outputs from a 2D spherical set-up with different initial molecular cloud core masses and from a 1~$M_{\odot}$ 2D simulation that includes an initial angular momentum (dashed blue line). Each plot includes a top subplot with a linear \facc axis ranging from 0 to 1. Typical constant values of \facc used in the literature lie within the grey hatched region.}
\label{fig:2DAvg-outsideshock}
\end{figure}

For completeness and comparison, in this section, we briefly present results from 2D \texttt{PLUTO} simulations with and without rotation. 

We follow the protostellar evolution for a few tens of years after its formation in the 2D spherically symmetric runs. Due to computational time restrictions for the 2D \texttt{PLUTO} run with rotation, we simulate the protostellar evolution for $\sim$59~years but here for comparison show results only for the initial three years before an outflow is launched.

In the case of the simulation including an initial angular momentum, the accretion shock overlays on both the protostar and its surrounding disc. This makes it challenging to decouple the two and define the protostellar radius as the accretion shock position \citep[as shown in][]{Bhandare2024, Ahmad2024}. Hence for this \texttt{PLUTO} run, we use the midplane $\mathrm{H_2}$ dissociation front as an estimate for the second core radius. This ensures choosing a radius at which most $\mathrm{H_2}$ is dissociated, a condition that leads to protostellar birth. The radiative efficiency \facc and the accretion rate shown in Fig.~\ref{fig:2DAvg-outsideshock} are then computed just outside of this location. 

The time-varying behaviour of \facc rapidly approaching unity and a decreasing accretion rate in the 2D \texttt{PLUTO} data are very similar to that seen in the 1D \texttt{PLUTO} simulations.

\section{Complementary measurements pertaining to \facc}

\subsection{Time evolution of the accretion and radiative fluxes}
\label{appendix:localfacc}

\begin{figure}[!tp]
\centering
\begin{subfigure}{0.45\textwidth}
\includegraphics[width=0.9\textwidth]{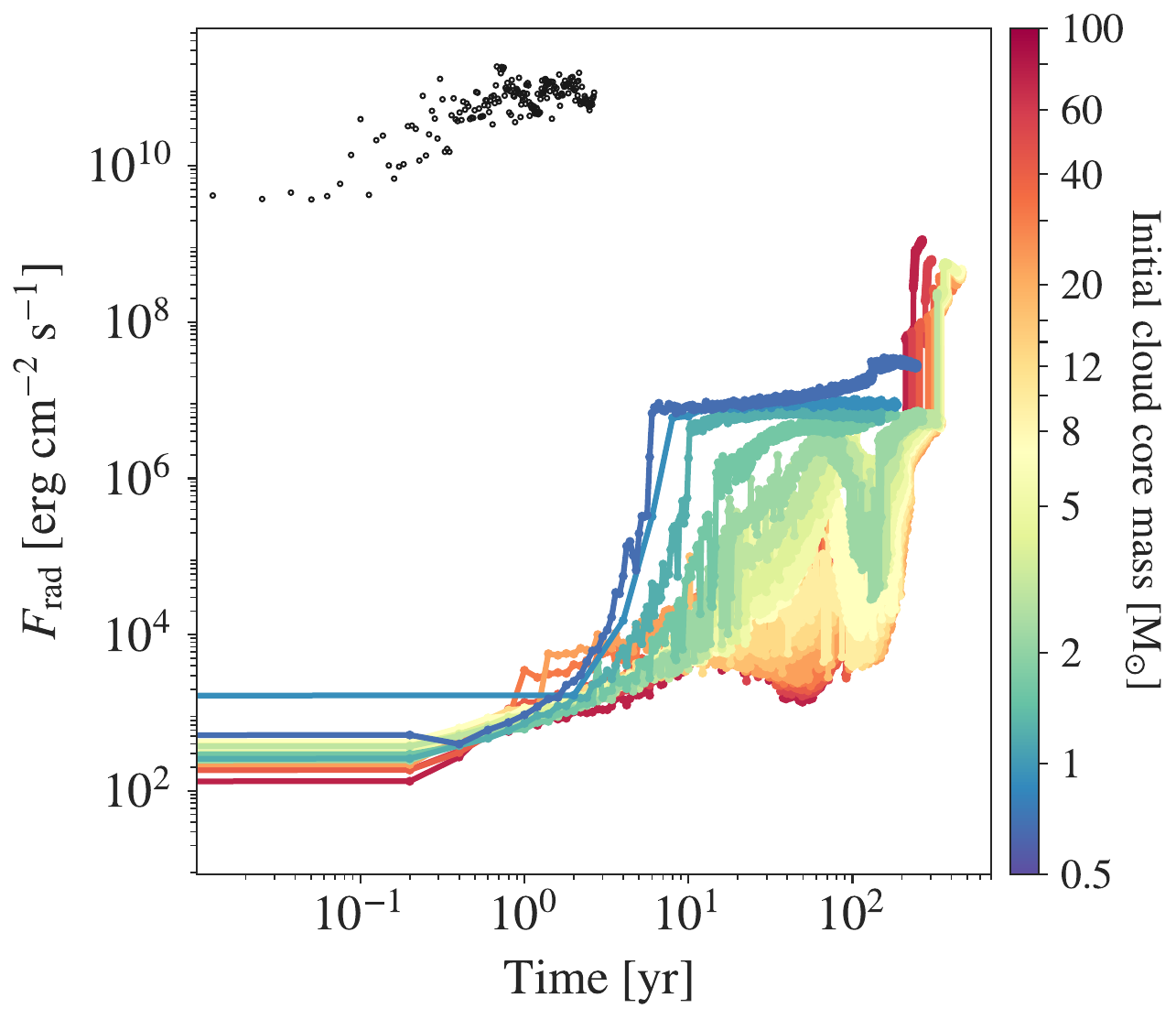}
\vspace{0.1cm}		
\end{subfigure}
\begin{subfigure}{0.45\textwidth} \includegraphics[width=0.9\linewidth]{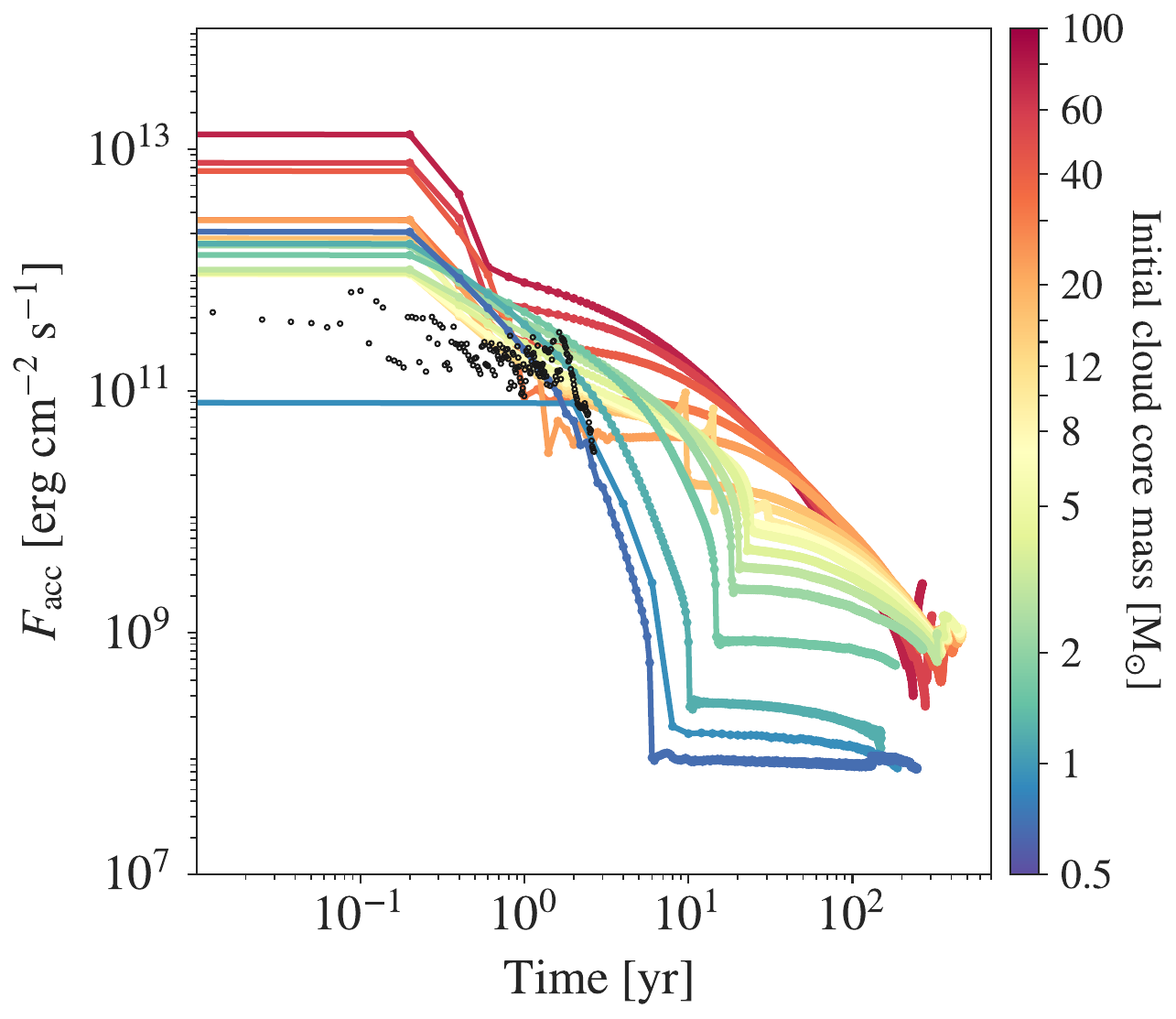}
\end{subfigure}
\caption{Time evolution of the radiative flux $F_{\mathrm{rad}}$ (\textit{top}) and accretion energy flux $F_{\mathrm{acc}}$ (\textit{bottom}) since protostellar birth. Both panels show outcomes from a 1D spherical set-up with different initial molecular cloud core masses using the \texttt{PLUTO} code. Also shown as black circles in both panels is data from a 3D \texttt{RAMSES} simulation with an initial turbulent Mach number of 0.2.}
\label{fig:FaccFrad}
\end{figure}

In this letter, we have shown that \facc substantially increases over time. It is of interest to assess whether this increase is due to a drop in the incident accretion energy flux $F_{\mathrm{acc}}$, or an increase in the outgoing radiative flux $F_{\mathrm{rad}}$. To this end, in Fig.~\ref{fig:FaccFrad}, we present the time evolution of $F_{\mathrm{rad}}$ (top) and $F_{\mathrm{acc}}$ (bottom) from our 1D \texttt{PLUTO} runs. We see that as time progresses, $F_{\mathrm{acc}}$ decreases due to the steady reduction in mass accretion rate; however, $F_{\mathrm{rad}}$ simultaneously increases as the protostellar shock front continuously expands to lower density (and hence optically thinner) regions. By the end of each simulation, the simultaneous increase in $F_{\mathrm{rad}}$ and drop in $F_{\mathrm{acc}}$ causes the shock front to reach \facc $\approx 1$, thus achieving supercriticality.

In the 3D run however, accretion onto the nascent protostar is anisotropic (as discussed in the main body of this letter). As such, the rise in \facc is instead explained by a rapid increase in $F_{\mathrm{rad}}$ as soon as a polar density cavity is formed, thus providing the means for radiation to escape. This is shown in the sub-section below (Appendix~\ref{appendix:globalfacc}).

\subsection{Global measurement of \facc}
\label{appendix:globalfacc}

\begin{figure}[!tp]
\begin{subfigure}{0.38\textwidth}
\includegraphics[width=\textwidth]{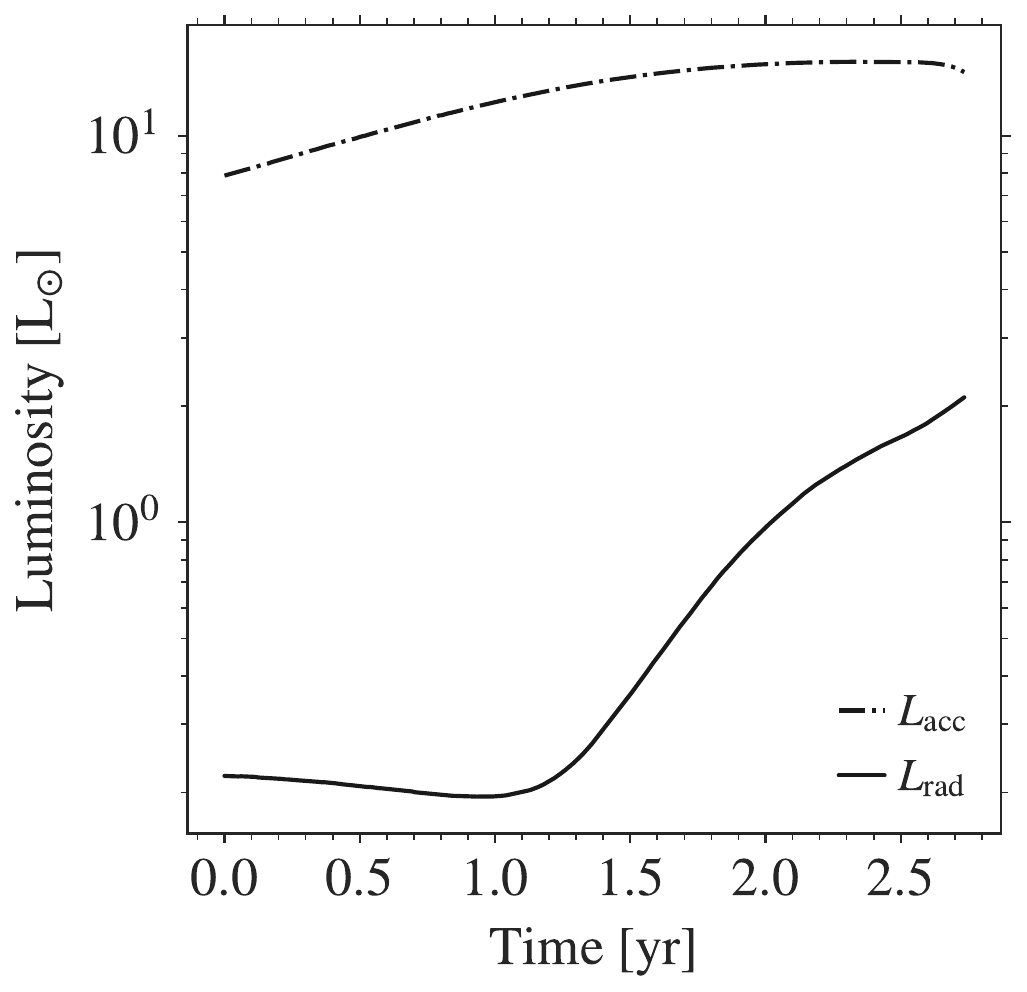}
\vspace{0.1cm}		
\end{subfigure}
\begin{subfigure}{0.45\textwidth} \includegraphics[width=\linewidth]{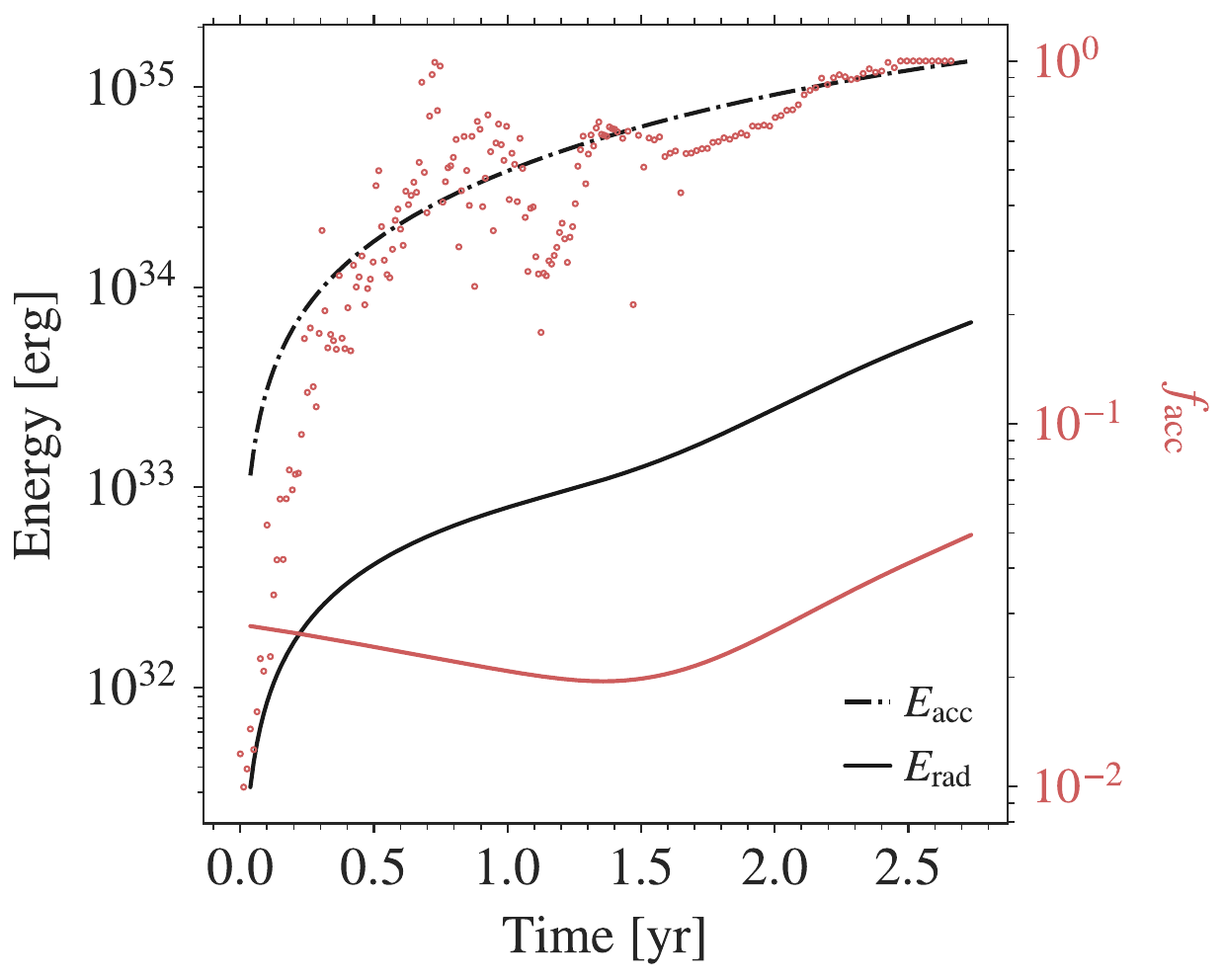}
\end{subfigure}
\caption{Global measurement of the radiative efficiency of the 3D \texttt{RAMSES} run as a function of time (where $t=0$ corresponds to the epoch of protostellar birth), performed at $R_{\mathrm{s}}=1.2$ au (see Eq.~\ref{eq:faccglobal}). Top panel displays the outgoing radiative luminosity (solid line) and accretion luminosity (dash-dotted line). Bottom panel presents the time integrated luminosities, yielding the radiative energy (solid line) and the accretion energy (dash-dotted line). Bottom panel also shows the global \gfacc (red line) and local (red circles) radiative efficiency.}
\label{fig:faccenergy}
\end{figure}

In addition to the local and instantaneous measurements of \facc presented in this letter, it is of interest to measure this parameter from an energetic point of view. For this, we extract a spherical shell with a fixed radius $R_{\mathrm{s}}$, encompassing both the protostar and disc, where we measure $F_{\mathrm{acc}}$ and $F_{\mathrm{rad}}$, which we then use to compute the total energy crossing into and out of the shell
\begin{equation}
\mathrm{^{global}}f_{\mathrm{acc}}(t) = \frac{\int_{0}^{t}\int_{S}F_{\mathrm{rad}}dSdt}{\int_{0}^{t}\int_{S}F_{\mathrm{acc}}dSdt} = \frac{\int_{0}^{t}L_{\mathrm{rad}}dt}{\int_{0}^{t}L_{\mathrm{acc}}dt} = \frac{E_{\mathrm{rad}}}{E_{\mathrm{acc}}},
\label{eq:faccglobal}
\end{equation}
where $S$ is the surface of the spherical shell of radius $R_{\mathrm{s}}$, $F_{\mathrm{rad}}$ is the outgoing radiative flux defined in Eq.~\ref{eq:flux}, and $F_{\mathrm{acc}}$ is the flux of incoming accretion energy defined in Eq.~\ref{eq:accflux}. $L_{\mathrm{rad}}$ and $L_{\mathrm{acc}}$ are the surface integrated radiative and accretion luminosities, respectively. $E_{\mathrm{rad}}$ and $E_{\mathrm{acc}}$ are the time integrated radiative and accretion energy, respectively.

The resulting measurements of \gfacc at $R_{\mathrm{s}}=$~1.2~au are presented in the bottom panel in Fig.~\ref{fig:faccenergy} (red curve). We see that in the 3D \texttt{RAMSES} run, the outgoing radiative energy has yet to match the incoming accretion energy during the short simulated timespan ($\sim$~2.5 years). Nevertheless, when the protostellar shock front reaches local supercriticality after one year (i.e., the energy radiated just upstream of the shock front equals the accretion luminosity, see Fig.~\ref{fig:Avg-outsideshock}), it begins to radiate a substantial amount of energy, driving $L_{\mathrm{rad}}$ up and causing the global radiative efficiency to increase substantially. Given enough evolution time, even the global radiative efficiency will reach unity.

\end{appendix}

\end{document}